\documentclass{PoS}
\usepackage{epsfig}

\title{Charm spectroscopy on dynamical 2+1 flavor domain
       wall fermion lattices with a relativistic heavy
       quark action}

\ShortTitle{Charm spectroscopy on dynamical 2+1f DWF lattices with a RHQ action}

\author{\speaker{Min Li}\\%
        Department of Physics, Columbia University, New York, NY 10027, USA\\
        E-mail: \email{minxolee@phys.columbia.edu}}

\author{Huey-Wen Lin\\
        Thomas Jefferson National Accelerator Facility, Newport News, VA 23606, USA\\
        E-mail: \email{hwlin@jlab.org}}

\author{RBC and UKQCD collaborations}

\abstract{We present a preliminary calculation of the charmonium spectrum using the dynamical 2+1 flavor $24^3\times 64$ domain wall fermion lattice configurations generated by the RBC and UKQCD collaborations. We use the relativistic heavy quark action with 3 parameters non-perturbatively determined by matching to experimental quantities. Chiral extrapolation is done on four light sea quark masses from 0.005 to 0.03, with $m_s$=0.04 and $m_{res}$=0.003. We can either predict meson masses assuming the lattice spacing is known from other methods, or calculate the lattice spacing using those quantities.}

\FullConference{The XXV International Symposium on Lattice Field Theory\\
                 July 30 - August 4 2007\\
                 Regensburg, Germany}

\begin{document}

\section{Introduction}
   Flavor physics is one of the urgent applications of Lattice QCD. However, the fact that the heavy quark masses are large in lattice units is a long-standing problem for heavy quark physics with LQCD. In the application, $ma\ll 1$ is no longer true and the terms containing $(ma)^n$ (with $a$ the lattice spacing) become significant. As a direct simulation with $a \ll 1/m$ costs too much, we resort to effective field theories.  Various heavy quark effective actions were developed and used for different physical systems, see Refs. ~\cite{Kronfeld:2003sd,Wingate:2004xa,Okamoto:2005zg,Onogi:2006km} for reviews on this topic.
   
   In this proceeding, our work is based on the so called relativistic heavy quark (RHQ) action~\cite{ElKhadra:1996mp, Aoki:2001ra,Christ:2006us,Christ:lat06}. The lattice form of the action, following the
formulation proposed in ~\cite{Christ:2006us,Christ:lat06}, can be written as:
   \begin{equation}
   S = \sum_n \overline{\Psi}_n\left\{ m_0 + \gamma_0D_0 -\frac{1}{2}aD^2_0 + \zeta\left[ \vec{\gamma}\cdot\vec{D}-\frac{1}{2}a(\vec{D})^2\right] -a\sum_{\mu\nu} \frac{i}{4}c_P\sigma_{\mu\nu}F_{\mu\nu}\right\} \Psi_n
   \end{equation}
   In the heavy quark case, the temporal covariant derivative $D_0$ is around the order of $ma$ and should not be treated the same way as the spatial derivatives $D_i$, which are of order $\Lambda_{QCD}a$ or $\alpha_sma$ depending on the system under investigation. Following the Symanzik improvement procedure, we found that only the three free parameters $m_0$, $c_P$ and $\zeta$ need to be tuned to remove all errors of order $(m a)^n$ and $|\vec p a|$.  Thus, if the parameters are correctly tuned, the action will have small cutoff effects: $(\Lambda_{QCD}a)^2$ for heavy-light systems and $(\alpha_sma)^2$ for heavy quarkonium. The main purpose of this work is to determine the three parameters by matching to physical quantities for charmed systems, making more accurate predictions for charmed mesons possible. The lattice spacing can also be obtained with reasonable precision if we treat it as a fourth quantity to be adjusted to correctly predict the charmed meson spectrum. All work has been done on dynamical 2+1 flavor lattices, which is a continuation of work done by H.-W. Lin~\cite{Lin:2007qu}.
   
   The lattices used in this work are the dynamical 2+1 flavor $24^3\times 64$ DWF lattice configurations generated by the RBC and UKQCD collaborations~\cite{MfEnno:2007lat}. For each configuration, we place sources at times 0, 16, 32 and 48 separately for better statistics; see Fig.~\ref{fig:a1pion}. Part of the data was collected and the analysis was done during and after the lattice conference. This additional data is included in this proceeding for completeness. Binning the data every two configurations had no effects on the results, which suggests the auto-correlation of the lattice configurations is negligible. 
   
   \begin{table}[ht]
      \centering
      \begin{tabular}{ccccc}
	\hline
	\hline
	volume & $L_s$ & ($m_{sea},m_s$) &  Traj(step)& \# of configs \\
	\hline
	$24^3 \times 64$  & 16 & (0.005,0.04) & 900-4500(40) & 91   \\
	$24^3 \times 64$  & 16 & (0.01,0.04)  & 900-4500(40) &  91    \\
	$24^3 \times 64$  & 16 & (0.02,0.04)  & 1885-3605(20) & 87   \\
	$24^3 \times 64$  & 16 & (0.03,0.04)  & 1000-3060(20) & 104  \\
	\hline
      \end{tabular}
      \label{tab:lattices}
   \end{table}
   
\section{Determine the RHQ action and the lattice spacing}
   To determine the action in such a way that errors are  controllable, we tune the parameters by matching physical observables sensitive to them to their experimental values. The parameters are then determined for each ensemble with different light sea quark masses and extrapolated to the chiral limit. The physical on-shell quantities we are going to use are mass combinations of pseudo-scalar (PS), vector (V), scalar (S) and axial-vector (AV) mesons in heavy-heavy (hh) and heavy-light (hl) systems~\cite{Lin:2006ur}. 
   \begin{itemize}
   \item spin-averaged:   $m^{hh}_{sa} = \frac{1}{4}(m^{hh}_{PS} + 3m^{hh}_V)$,  
      $m^{hl}_{sa} = \frac{1}{4}(m^{hl}_{PS} + 3m^{hl}_V)$  
   \item hyperfine splitting:  $m^{hh}_{hs} = m^{hh}_V - m^{hh}_{PS}$, 
      $m^{hl}_{hs} = m^{hl}_V - m^{hl}_{PS}$
    \item mass ratio:  $\frac{m_1}{m_2}$, where $E^2 = m_1^2 + \frac{m_1}{m_2}p^2$, $m_1$: rest mass, $m_2$: kinetic mass. 
    \item spin-orbit averaged and splitting: $m^{hh}_{sos} = m^{hh}_{AV} - m^{hh}_S$,
      $m^{hh}_{soa} = \frac{1}{4}(m^{hh}_{S} + 3m^{hh}_{AV})$
   \end{itemize}
  
  With the experimental values of these quantities at hand, we use a linear ansatz relating the three parameters ($X_{RHQ}$) and the corresponding measured quantities ($Y(a)$). The linear approximation only  holds in a limited region of the parameter space, which we estimate from earlier, dynamical $16^3\times 32$ studies~\cite{Lin:2007qu}.
  \vspace{-0.2cm}
        
   \begin{equation}
	Y(a) = \left(
	\begin{array}{c}
	  m_{\eta_c} a \\
	  m_{J/\psi} a \\
	  ...\\
	  ... \\
	  m_1/m_2 
	\end{array} 
	\right) = J\cdot X_{RHQ}=J\cdot \left( 
	\begin{array}{c}
	  m_0a \\
	  c_P \\
	  \zeta
	\end{array} 
	\right)
	+ A  \ ,\nonumber
    \end{equation}
where the quantities $Y(a)$ are known if we assume the lattice spacing $a$ is known from another method or $a$-dependent if we treat $a$ as a free parameter to be determined. Provided we are able to determine the J matrix and A vector, we can obtain the parameters by minimizing $\chi^2$ defined as:
\begin{equation}
\chi^2 = (J\cdot X_{RHQ}+A-Y(a))^TW^{-1}(J\cdot X_{RHQ}+A-Y(a)) \ ,
\label{eqn:chi2}
\end{equation}
where $W$ is the correlation matrix estimated from the measured data. We choose to use only the diagonal part sometimes because the data might be too noisy to give a well-behaved $W$. The quantity $\chi^2$ is a quadratic function of vector $X_{RHQ}$ if lattice spacing $a$ is known and of the vector $(m_0a,c_P, \zeta, a)^T$ if $a$ is unknown, and so it is easy to minimize analytically. J and A can be calculated using finite differences directly from a Cartesian set, and in order to save time we collected data for the minimum (seven) number of parameter sets: centered at \{0.433,2.446,1.295\} and with extent \{0.1,0.1,0.02\}. There is a potential problem because the RHQ parameters which we finally determine are actually outside of the region bounded by the 7 sets of parameters which we studied. However, our earlier $16^3 \times 32$ work suggests the region of linearity extends to this matching point.
 
\section{Source search and other concerns}

\begin{figure}[ht]
    \vspace{-0.3cm}
    \hfill
    \begin{minipage}{0.45\textwidth}
        \hspace{-1.0cm}
        \epsfig{file=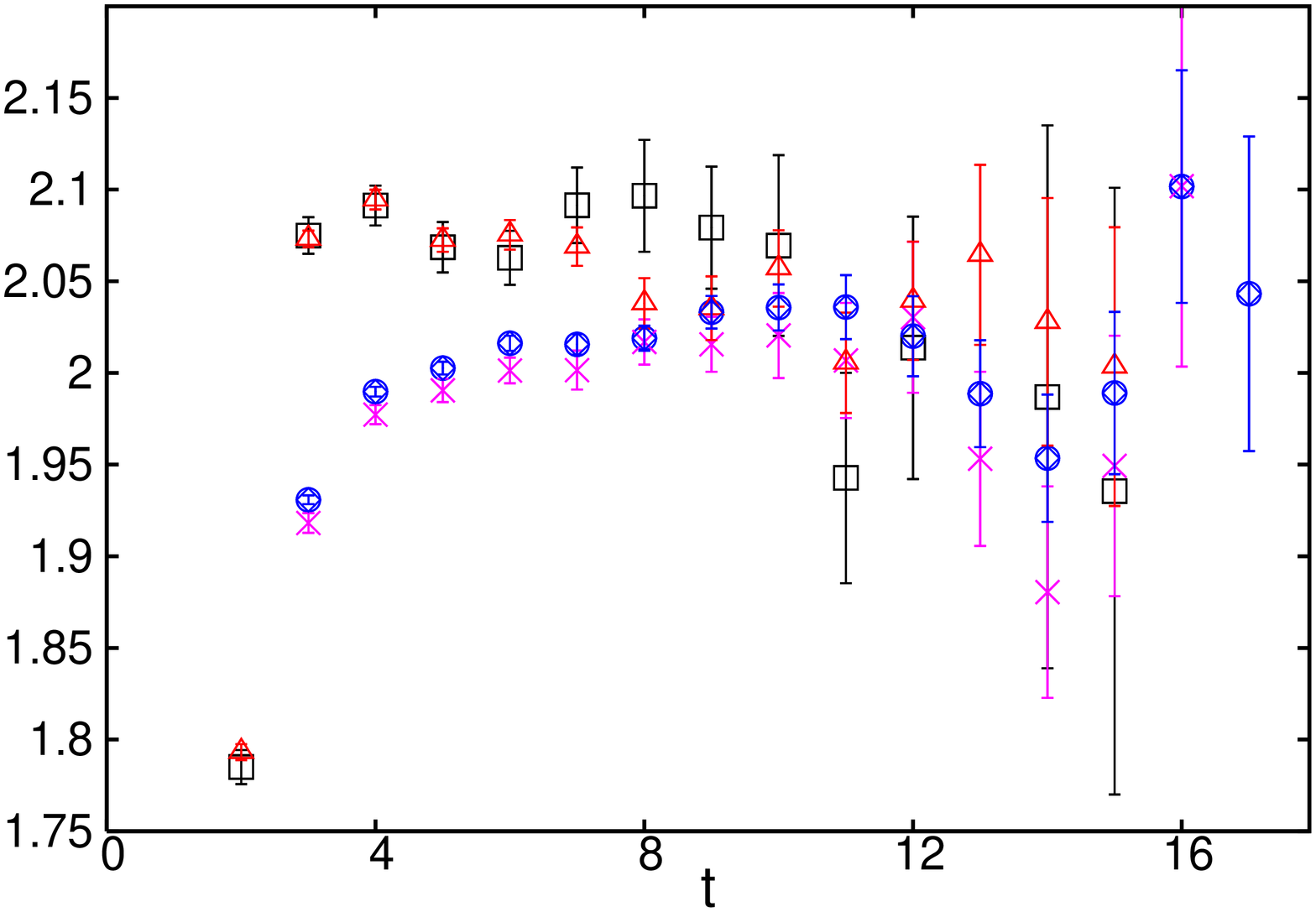,width=0.8\linewidth,angle=270} 
    \end{minipage}
    \hfill
    \begin{minipage}{0.45\textwidth}
        \hspace{-0.8cm}
        \epsfig{file=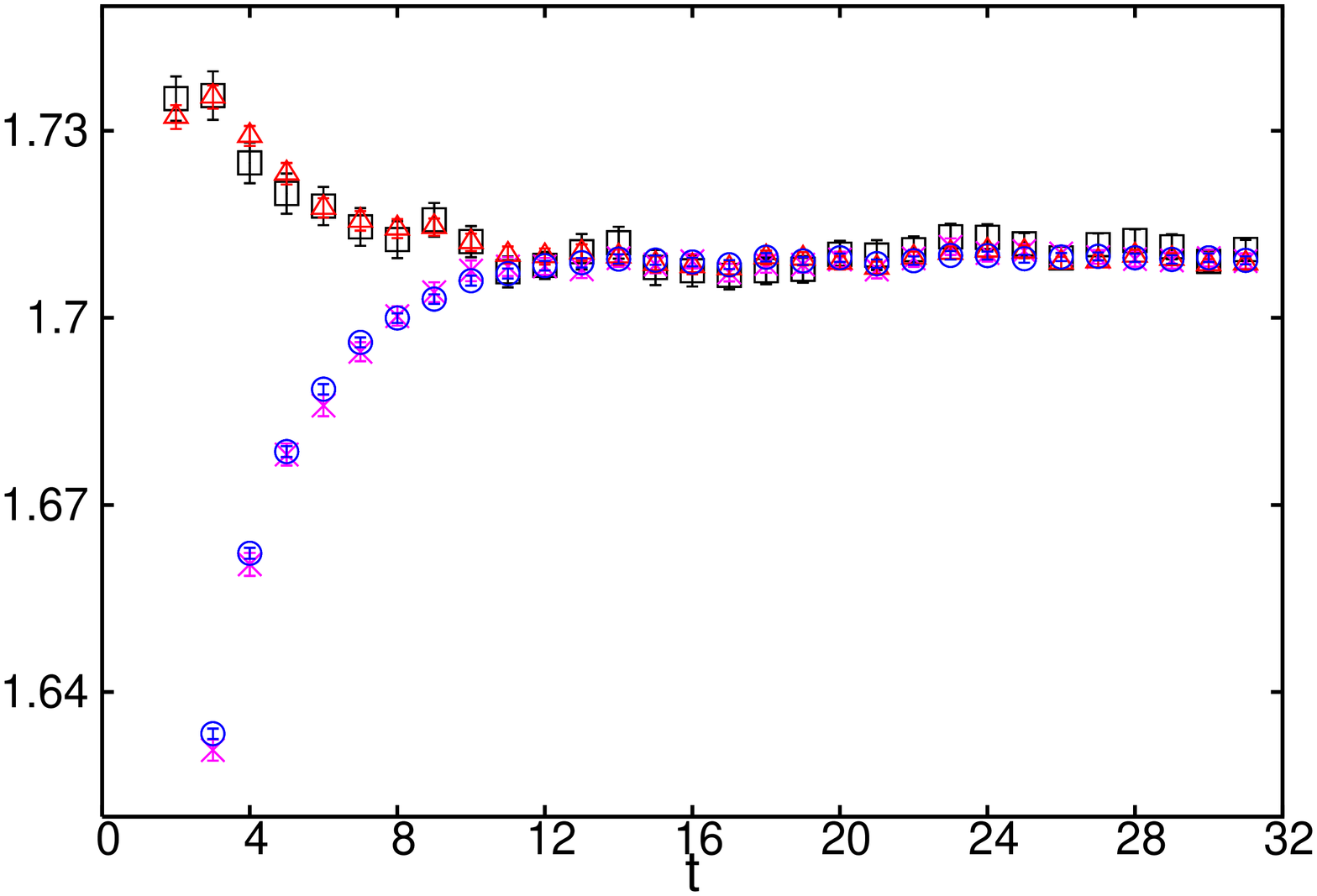,width=0.8\linewidth,angle=270}
    \end{minipage}
    \caption{Sample effective mass plots of $\chi_{c1}$ (left) and $\eta_c$ (right). Red triangles: 4 sources per config (s/c) and box size (bs) 4; blue circles: 4 s/c,  bs=12; black squares: 1 s/c, bs=4 and magenta crosses: 1 s/c, bs=12.}
    \label{fig:a1pion}
    \hfill
    \vspace{-0.4cm}
\end{figure}
 
After studying a number of box sources we found that sources with box size 4 and 12 are the best to extract masses of the pseudo-scalar $\eta_c$ and vector $J/\psi$ using a two state fit ($t\in[4,24]$), Fig.~\ref{fig:a1pion} (right). However, for masses of the scalar $\chi_{c0}$ and axial-vector $\chi_{c1}$, the effective mass plot, Fig.~\ref{fig:a1pion} (left) shows that without enough statistics the box size 4 source tends to give a false plateau, so we use the box size 12 source to determine the $\chi_{c0}$ and $\chi_{c1}$ masses via a single state fit ($t\in[9,15]$). Currently the heavy-strange data use a box source with size 8, aiming for the best plateau. The $D_s$ and $D_s^*$ states are using fitting ranges $t\in[6,32]$ and $[12,32]$ respectively. For the mass ratio $m_1/m_2$, the momentum dependence is studied for both the $\eta_c$ and $J/\psi$ mesons and the results are quite consistent. We use results from the $\eta_c$ momentum dependence with the three lowest momenta. 

Other concerns such as quark propagator inversion precision are studied in detail for heavy quarks and the relative error for every time slice is controlled to less then $10^{-4}$ when a source placed at time zero, while light propagators are well-understood from previous studies.

\section{Analysis and results}
Let's list explicitly all the quantities used here for fitting, (1)$\frac{1}{4}(m_{\eta_c}+3m_{J/\psi})$ (2) $m_{J/\psi}-m_{\eta_c}$ (3) $\frac{m_1}{m_2}$ (4) $\frac{1}{4}(m_{\chi_{c0}}+3m_{\chi_{c1}})$ (5) $m_{\chi_{c1}}-m_{\chi_{c0}}$ (6) $\frac{1}{4}(m_{D_s}+3m_{D_s^*})$ (7) $m_{D_s^*}-m_{D_s}$ 

\subsection{Heavy-heavy sector}
Using only results from the heavy-heavy states, i.e., from quantities (1)--(3), the matched RHQ parameters and the corresponding chiral extrapolation are shown in the table below:
\begin{table}[ht]
\begin{minipage}{0.45\textwidth}
      \begin{tabular}{|cccc|}
	\hline
	\hline
	$m_{sea}$ & $m_0a$ & $c_P$ & $\zeta$ \\ 
	\hline
	0.005 & 0.410(8)  & 2.356(16)  &  1.270(7)  \\
	\hline
	0.01  & 0.398(8) & 2.323(15)  &  1.269(9) \\
	\hline
	0.02  & 0.371(9)  & 2.263(14)  &  1.272(9) \\
	\hline
	0.03  & 0.341(7) & 2.170(14)  &  1.263(8) \\
	\hline
	\hline
	-0.00315 &  0.434(9) & 2.422(18) &  1.273(9)\\
	\hline
	\hline 
     \end{tabular}
\end{minipage}	
\begin{minipage}{0.45\textwidth}
      \hspace{0.7cm}
      \begin{tabular}{|cccc|}
	\hline
	\hline
	$m_{sea}$ & $m_0a$ & $c_P$ & $\zeta$ \\ 
	\hline
	0.005 & 0.228(9)  & 2.029(15)  &  1.238(8)  \\
	\hline
	0.01  & 0.217(8) & 1.998(15)  &  1.237(9) \\
	\hline
	0.02  & 0.190(9)  & 1.940(13)  &  1.240(10) \\
	\hline
	0.03  & 0.162(8) & 1.853(13)  &  1.230(8) \\
	\hline
	\hline
	-0.00315&  0.251(9) & 2.091(17) &  1.242(10)\\
	\hline
	\hline	
      \end{tabular}
 \end{minipage}
   \caption{The inverse lattice spacing is assumed to be 1.62 GeV (from the static quark potential with $r_0=0.50fm$) for the left table  and 1.73 GeV (from $\Omega^-$ baryon) for the right one.}
    \label{tab:RHQparas}
\end{table}

\begin{figure}[ht]       
    \vspace{-0.3cm}
    \hfill
    \begin{minipage}{0.45\textwidth}
    	\hspace{-1.0cm}
          \epsfig{file=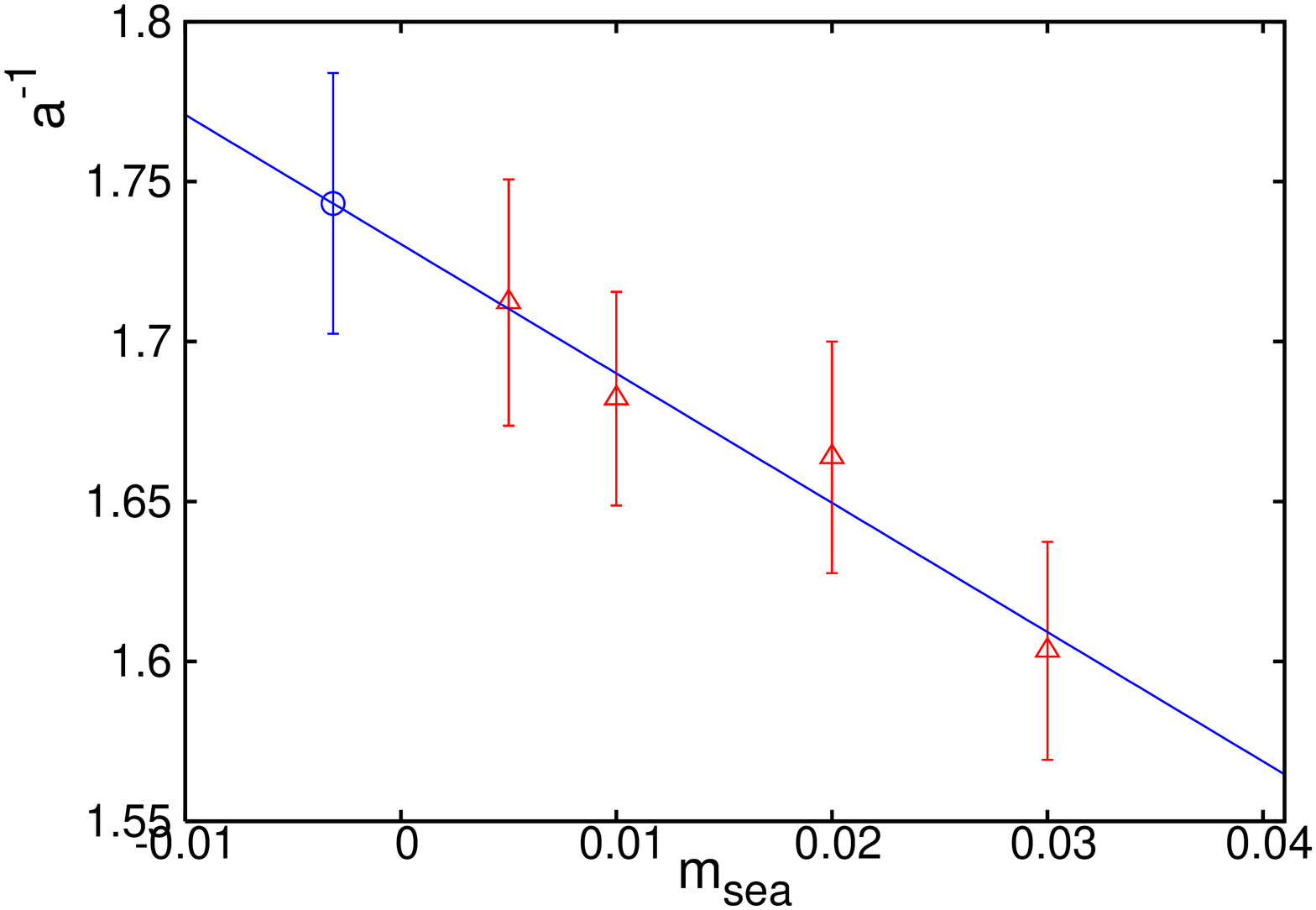,width=0.8\linewidth,angle=270}
          \caption{Chiral extrapolation of inverse lattice spacing, determined from quantities (1) to (5), with $W$ diagonal correlation matrix}  
           \label{fig:ainvChiraldiag-noDs}
    \end{minipage}
    \hfill
    \begin{minipage}{0.45\textwidth}
         \hspace{-0.7cm}
         \epsfig{file=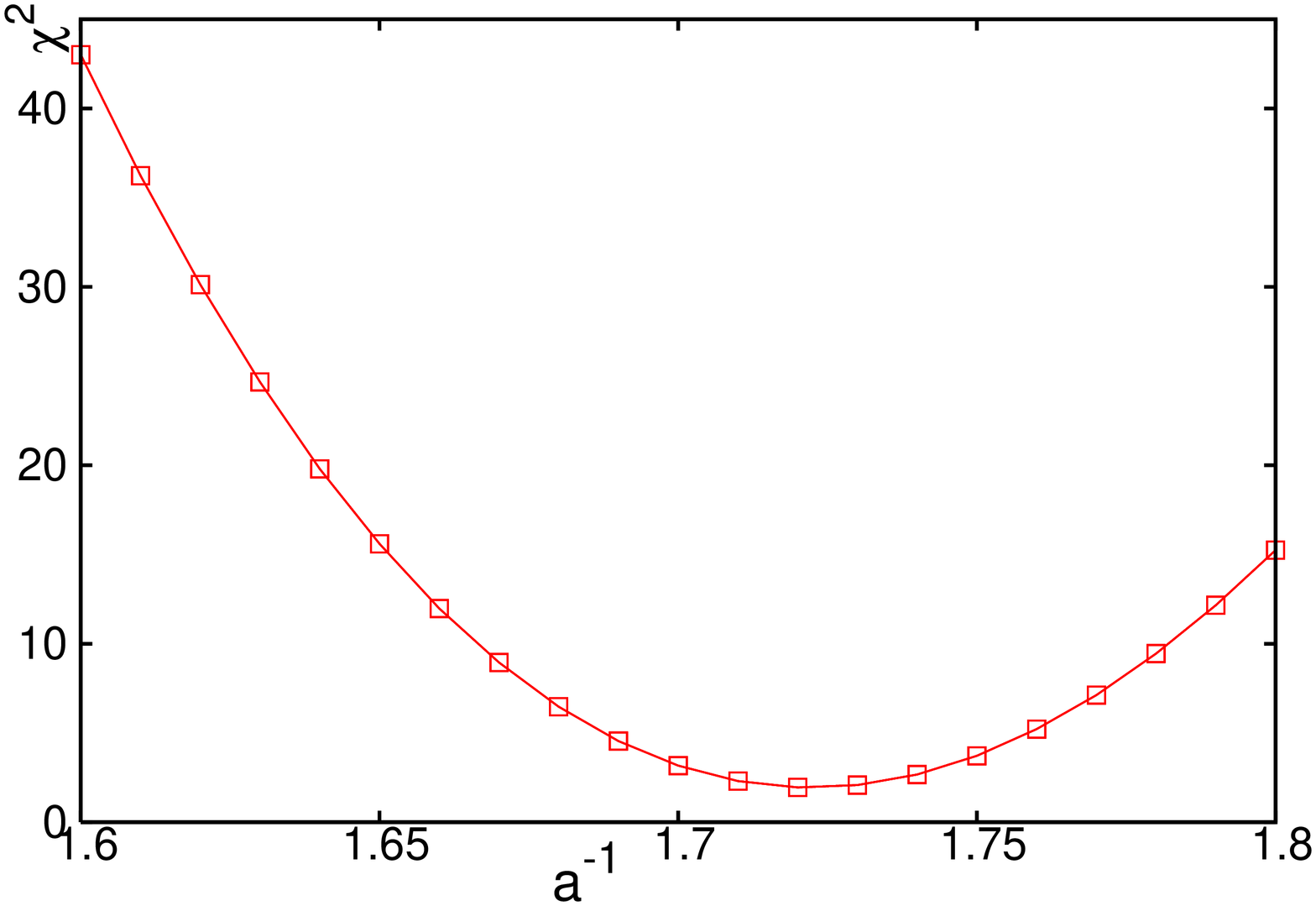,width=0.8\linewidth,angle=270}	
         \caption{The naive $\chi^2_{pred}$ at chiral limit  from fitting the three RHQ parameters with different input lattice spacings.}
      	 \label{fig:chi2-a}
    \end{minipage}
    \hfill
\end{figure}

From quantities (1)--(5), we can determine the RHQ parameters as well as the lattice spacing. Since the states $\chi_{c0}$ and $\chi_{c1}$ are a lot nosier the correlation matrix $W$ (in Eq.~\ref{eqn:chi2}) is not well measured. So instead we use only the diagonal part of the correlation matrix. The chiral extrapolation gives $a^{-1}=1.74(4)$ GeV, as shown in Fig.~\ref{fig:ainvChiraldiag-noDs}. The inconsistency between this and the result from the static quark potential suggests that $r_0$ is inaccurate.

\begin{figure}[ht]
    \hfill
    \begin{minipage}{0.45\textwidth}
         \hspace{-1.0cm}
        \epsfig{file=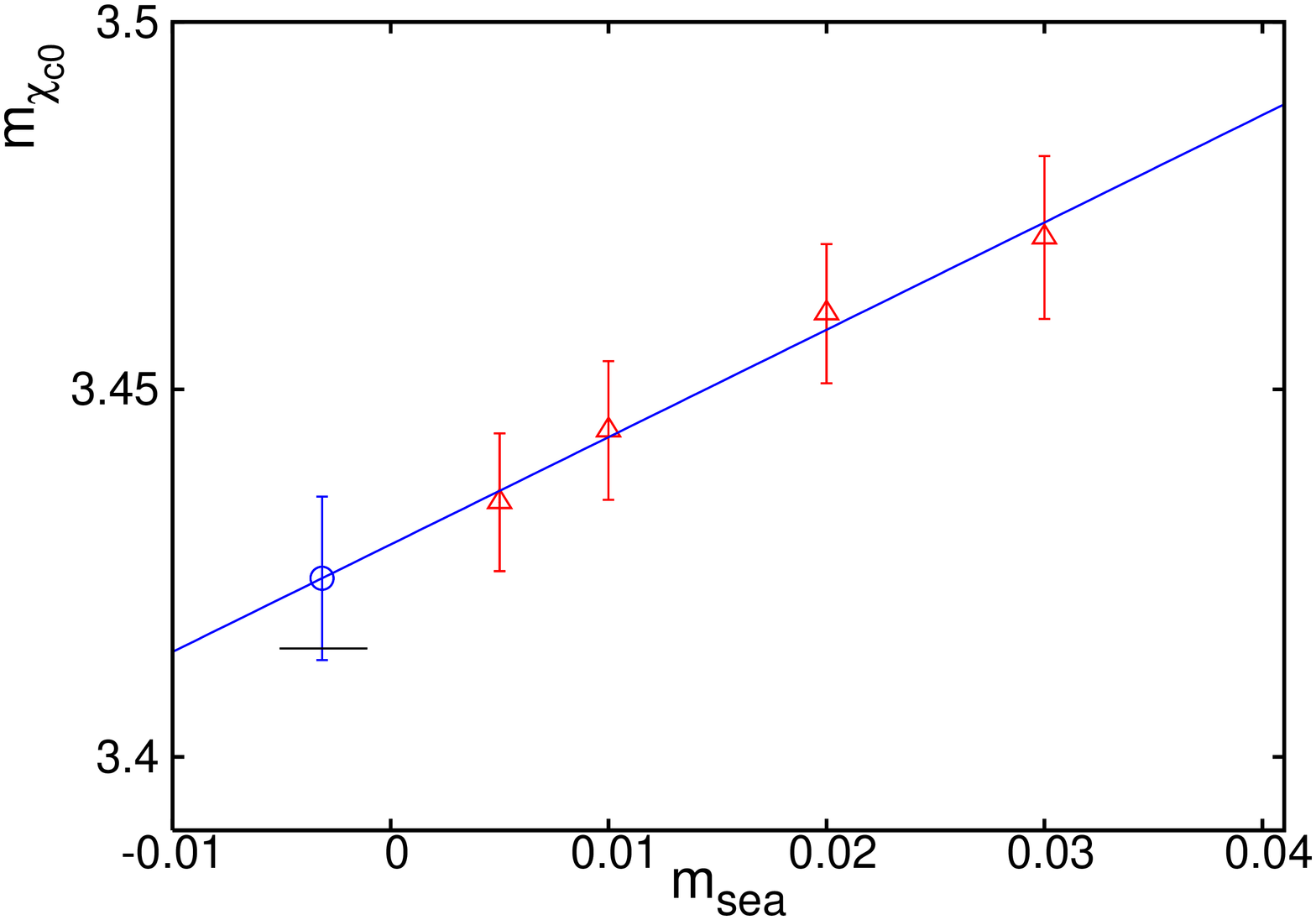,width=0.8\linewidth,angle=270} 
    \end{minipage}
    \hfill
    \begin{minipage}{0.45\textwidth}
        \hspace{-0.8cm}
        \epsfig{file=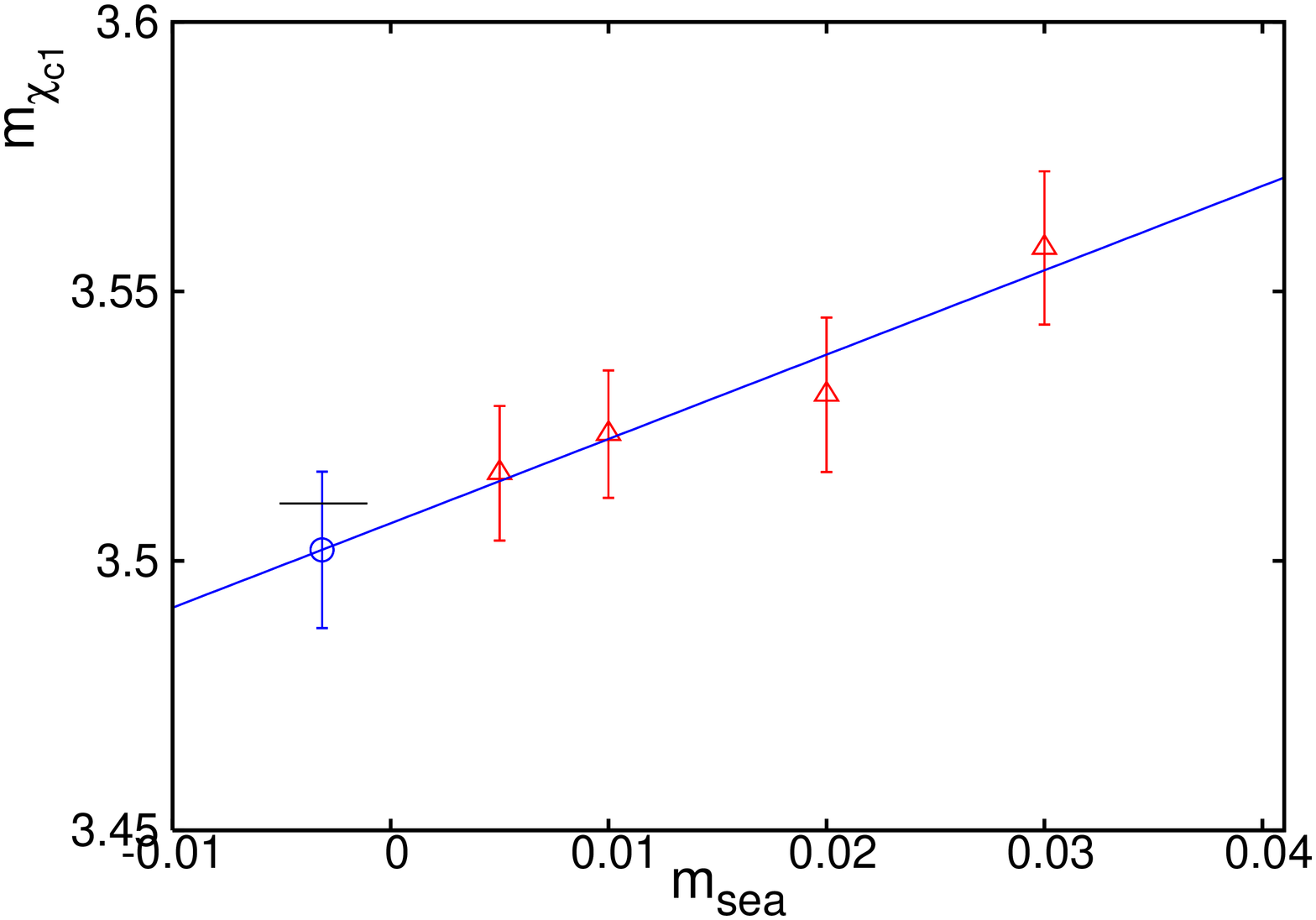,width=0.8\linewidth,angle=270}
    \end{minipage}
    \caption{The prediction of $m_{\chi_{c0}}$ (left) and $m_{\chi_{c1}}$ (right) in the chiral limit, with parameters determined above assuming $a^{-1}=1.73$ GeV, the black lines stand for experimental values.} 
    \label{fig:chic01-pred}
    \hfill
    \vspace{-0.4cm}
\end{figure}

We can make predictions for $\chi_{c0}$ and $\chi_{c1}$ states by using J and A calculated from the measured data and the RHQ parameters determined from quantities (1)--(3). See Fig.~\ref{fig:chic01-pred}; all errors are propagated using the jackknife technique. When extrapolated to the chiral limit, our predictions are quite consistent with the experimental values, and the errors are less than one percent. This is encouraging and suggests that we may apply this method to make accurate predictions for other charmed mesons. 

Some consistency checks have been carried out for the fitting procedure, especially for the fitting which determines the lattice spacing. We treat lattice spacing as an input parameter, and fit the RHQ parameters with predictions of $\chi_{c0}$ and $\chi_{c1}$ extrapolated to the chiral limit as a function of $a$. Then an uncorrelated, naive $\chi^2_{pred}$ is defined from:
\begin{equation}
\chi^2_{pred} = \sum_{i=0,1}\frac{ (m^{pred}_{\chi_{ci}}-m^{phys}_{\chi_{ci}})^2 }{\sigma^2(m^{pred}_{\chi_{ci}})}
\end{equation}
Here $m^{phys}_{\chi_{ci}}$ means the experimental value for the $\chi_{\chi_{ci}}$ meson. A plot showing the resulting $\chi^2_{pred}$ versus the inverse lattice spacing is plotted in Fig.~\ref{fig:chi2-a}. It shows good consistency that the $\chi^2_{pred}$ minimum occurs when $a^{-1}$ is around 1.72 GeV. For all fitting procedures, such as mass fitting and momentum dependence fitting, we use an uncorrelated fit.

\subsection{Heavy-strange sector}
\begin{figure}[ht]
    \vspace{-0.3cm}
    \hfill
    \begin{minipage}{0.45\textwidth}
        \hspace{-1.0cm}
        \epsfig{file=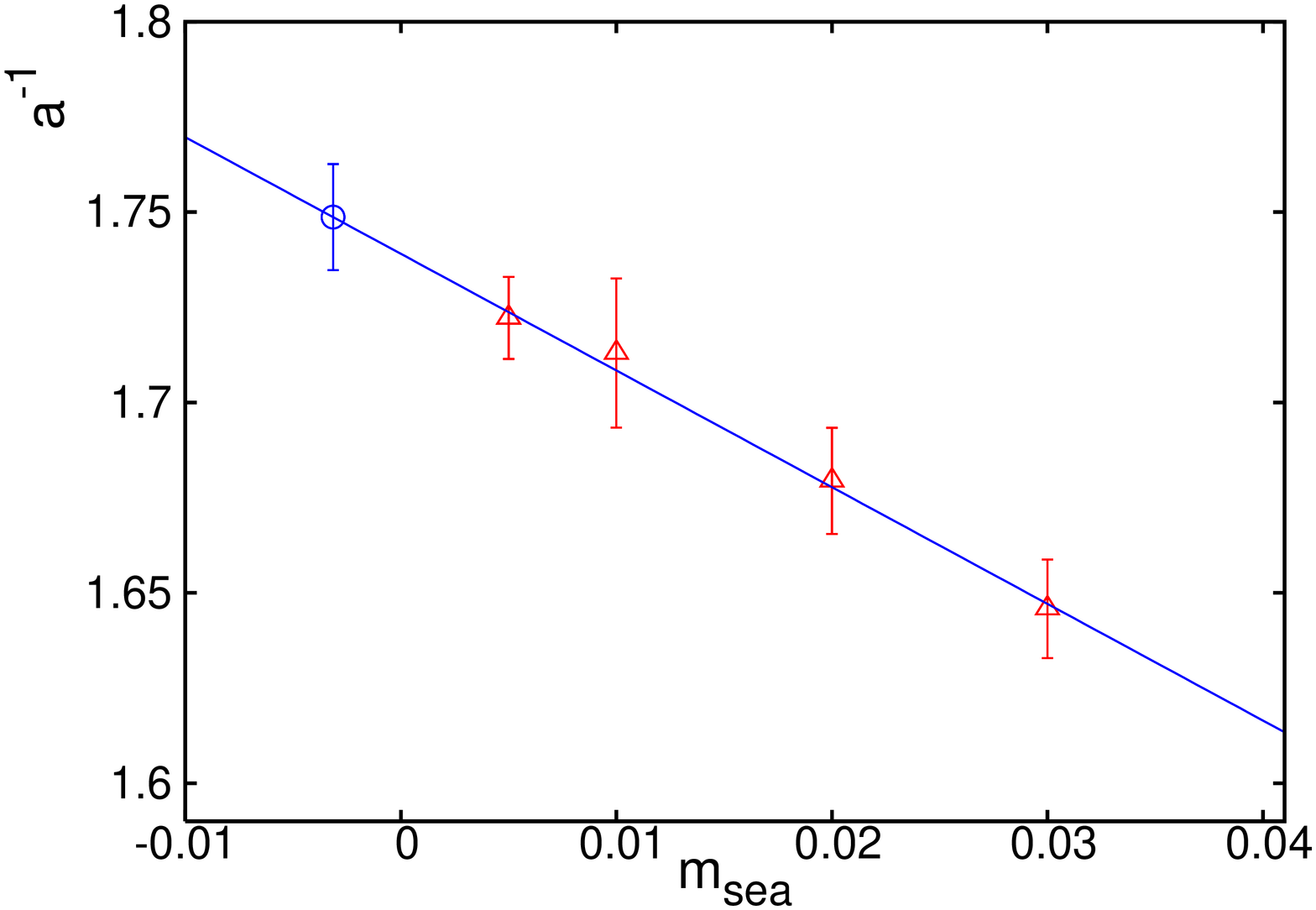,width=0.8\linewidth,angle=270} 
    \end{minipage}
    \hfill
    \begin{minipage}{0.45\textwidth}
        \hspace{-0.8cm}
        \epsfig{file=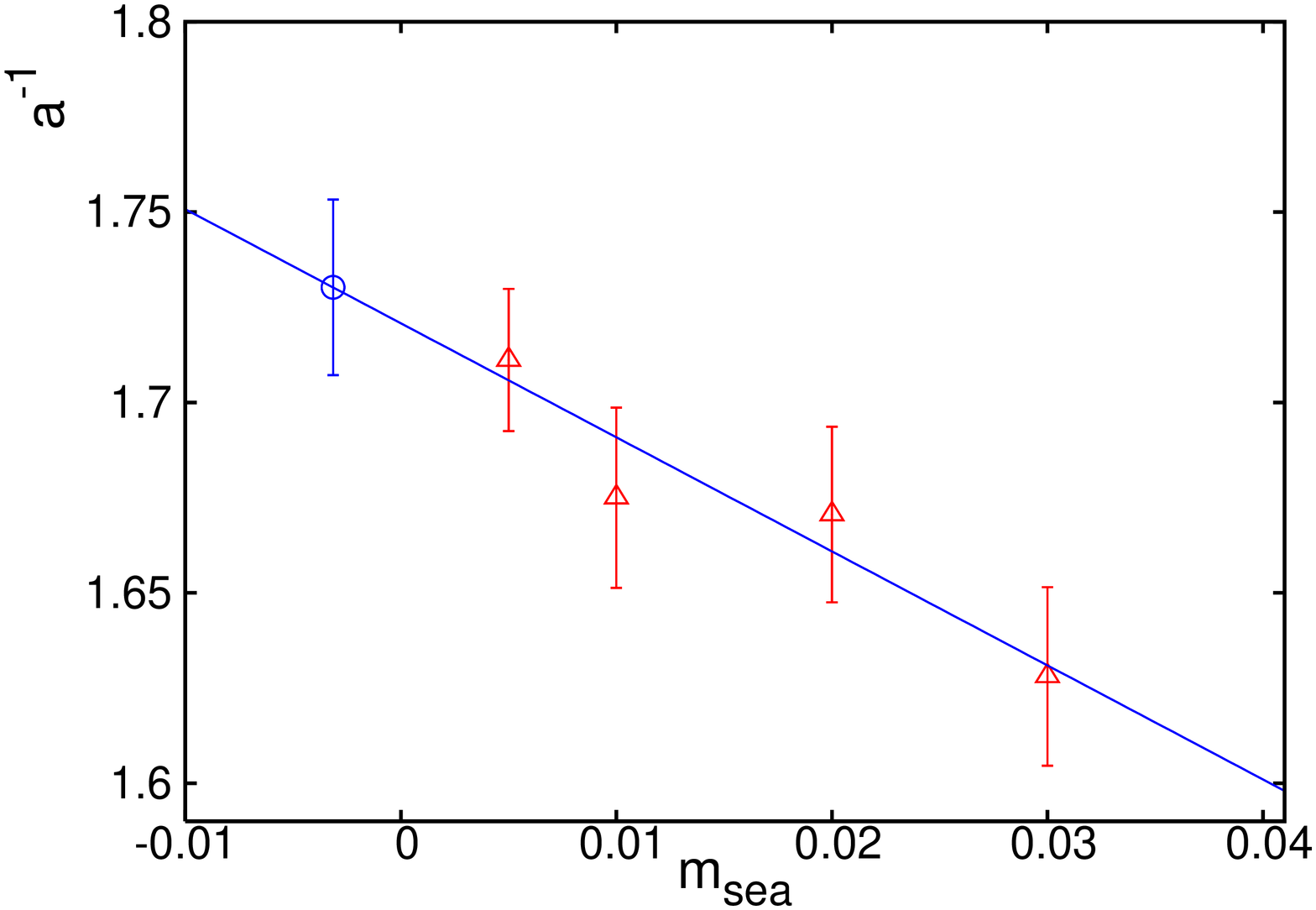,width=0.8\linewidth,angle=270}
    \end{minipage}
    \caption{Chiral extrapolation of inverse lattice spacing, determined from quantities (1)(2)(3)(6) and (7), with $W$ a full (left) or a diagonal (right) correlation matrix. } 
    \label{fig:ainv-wDs}
    \hfill
    \vspace{-0.2cm}
\end{figure}

The lattice spacing determined from chiral extrapolation above using heavy-heavy states is consistent with that determined from $\Omega^-$ baryon: $a^{-1}=1.73(2)$ GeV, but the errors are larger because of the noisy results for $\chi_{c0}$ and $\chi_{c1}$ states. So we proceed to include the heavy-strange sector, and include the quantities (6) and (7) in the analysis to replace (4) and (5) since (6) and (7) are more accurately determined. The physical strange quark mass we are using is $m_s=0.036$ in lattice units, Ref.~\cite{Allton:2007hx}; but as the $24^3$ data suggests a slightly different $m_s$, this may introduce some systematic error. We are now studying the more accurate value $m_s=0.034$, so we can extrapolate/interpolate to the right $m_s$ assuming the dependence on the strange mass is linear. The results of the fitted and chiral extrapolated RHQ parameters and $a$ with a full correlation matrix are listed below in Tab.~\ref{tab:latspacingDs}, and the extrapolations of $a^{-1}$ to the chiral limit are plotted for both cases with full correlation matrix and diagonal correlation matrix in Fig.~\ref{fig:ainv-wDs}. The corresponding results are 1.749(14) GeV and 1.730(23) GeV respectively.

\begin{table}[ht]
    \centering
    \begin{tabular}{ccccc}
      \hline
      \hline
      $m_{sea}$ & $m_0a$ & $c_P$ & $\zeta$ & $a^{-1}$(GeV)\\ 
      \hline
      0.005    & 0.241(21)   & 2.052(32)   &  1.240(8) &  1.722(11)    \\
      \hline
      0.01     & 0.243(38)   & 2.049(57)   &  1.242(9) &  1.713(20)    \\
      \hline
      0.02     & 0.271(30)   & 2.084(42)   &  1.254(9) &  1.679(14)     \\
      \hline
      0.03     & 0.297(27)   & 2.092(43)   &  1.254(8) &  1.646(13)    \\
      \hline
      \hline
      -0.00315 & 0.220(28)   & 2.037(42)   &  1.236(9) &  1.749(14)    \\
      \hline
      \hline
    \end{tabular}
    \caption{The RHQ parameters and lattice spacing determined from quantities (1)(2)(3)(6)(7), and extrapolated to the chiral limit, with $\chi^2$ from a full correlation matrix $W$.}
    \label{tab:latspacingDs}
\end{table}

 If we fix $a^{-1}$ to be 1.62 GeV then using (1)(2)(3)(6)(7) to determine the RHQ parameters will result in a huge $\chi^2/d.o.f=146/2$, which tells us the fitting fails if the wrong lattice spacing is used. If we set $a^{-1}$ to be 1.73 GeV, then $\chi^2/d.o.f=1.19/2$, which confirms again our observation.  

\section{Conclusion}
We have applied the RHQ action to the charmed system, both heavy-heavy and heavy-strange, and demonstrated that the parameters in the RHQ action can be determined with sub-percent precision by matching several quantities to their experimental values. We discovered our lattice spacing from static quark potential with $r_0=0.5fm$ was too large. Taking $a$ as a free parameter we were then able to determine it with a few percent error. The result is quite consistent with that implied by the $\Omega^-$ baryon.  In the heavy-heavy system, the $\chi_{c0}$ and $\chi_{c1}$ states are not as well-determined as the $J/\psi$ and $\eta_c$ states. We choose to use a diagonal correlation matrix when doing the four free parameters (three RHQ parameters and $a$) fitting in that case. The bare strange quark mass we are using in the heavy-strange run is $m_s=0.036$ in lattice units, which is slightly above the real one, so there might be a small amount of systematic error introduced into the results involving these states. Another heavy-strange run with different $m_s$ is underway. In conclusion, we view the application to the charmed system a success. And we will likely apply this method to charm-light states and perhaps to bottom quarks as well to explore more interesting topics.

\section*{Acknowledgment}
We acknowledge helpful discussions with Norman Christ and Robert Mawhinney. In addition, we thank Peter Boyle, Dong Chen, Norman Christ, Mike Clark, Saul Cohen, Calin Cristian, Zhi-hua Dong, Alan Gara, Andrew Jackson, Balint Joo, Chulwoo Jung, Richard Kenway, Changhoan Kim, Ludmila Levkova, Huey-Wen Lin, Xiaodong Liao, Guofeng Liu, Robert Mawhinney, Shigemi Ohta, Tilo Wettig and Azusa Yamaguchi for the development of QCDOC hardware and its software. The development and the resulting computer equipment were funded by the U.S. DOE grant DE-FG02-92ER40699, PPARC JIF grant PPA/J/S/1998/00756 and by RIKEN. This work was supported by U.S. DOE grant DE-FG02-92ER40699 and we thank RIKEN, BNL and the U.S. DOE for providing the facilities essential for this work.

\bibliographystyle{apsrev}
\bibliography{Proceeding}

\end{document}